\documentclass[aps,pre,showpacs,twocolumn]{revtex4}
\usepackage{graphics}
\usepackage{epsfig}

\begin{document}

\title{Topology of delocalization in the nonlinear Anderson model and anomalous diffusion on finite
clusters}
\author{A.V. Milovanov$^{1,2,4}$, A. Iomin$^{3,4}$}
\affiliation{$^1$ENEA National Laboratory,
Centro~Ricerche~Frascati, I-00044 Frascati, Rome, Italy \\
$^2$Space Research Institute, Russian Academy of Sciences, 117997
Moscow, Russia \\
$^3$Department of Physics and Solid State Institute, Technion,
Haifa 32000, Israel \\
$^4$Max-Planck-Institut f\"ur Physik komplexer Systeme, 01187
Dresden, Germany}

\begin{abstract}
This study is concerned with destruction of Anderson localization
by a nonlinearity of the power-law type. We suggest using a
nonlinear Schr\"odinger model with random potential on a lattice
that quadratic nonlinearity plays a dynamically very distinguished
role in that it is the only type of power nonlinearity permitting
an abrupt localization-delocalization transition with unlimited
spreading already at the delocalization border. For
super-quadratic nonlinearity the borderline spreading corresponds
to diffusion processes on finite clusters. We have proposed an
analytical method to predict and explain such transport processes.
Our method uses a topological approximation of the nonlinear
Anderson model and, if the exponent of the power nonlinearity is
either integer or half-integer, will yield the wanted value of the
transport exponent via a triangulation procedure in an Euclidean
mapping space. A kinetic picture of the transport arising from
these investigations uses a fractional extension of the diffusion
equation to fractional derivatives over the time, signifying
non-Markovian dynamics with algebraically decaying time
correlations.
\end{abstract}


\maketitle

\section{Introduction}

We consider the problem of dynamical localization of waves in a
nonlinear Schr\"odinger model with random potential on a lattice
and arbitrary power nonlinearity,
\begin{equation}
i\frac{\partial\psi_n}{\partial t} = \hat{H}_L\psi_n + \beta |\psi_n|^{2s} \psi_n,
\label{1s} 
\end{equation}
where $s$ ($s \geq 1$) is a real number;
\begin{equation}
\hat{H}_L\psi_n = \varepsilon_n\psi_n + V (\psi_{n+1} + \psi_{n-1})
\label{2} 
\end{equation}
is the Hamiltonian of a linear problem in the tight binding
approximation; $\beta$ ($\beta > 0$) characterizes the strength of
nonlinearity; on-site energies $\varepsilon_n$ are randomly
distributed with zero mean across a finite energy range; $V$ is
hopping matrix element; and the total probability is normalized to
$\sum_n |\psi_n|^2 = 1$. For $\beta\rightarrow 0$, the model in
Eqs.~(\ref{1s}) and~(\ref{2}) reduces to the original Anderson
model in Ref. \cite{And}. In the absence of randomness, the
nonlinear Schr\"odinger equation (NLSE) in Eq.~(\ref{1s}) is
completely integrable.

Experimentally, Anderson localization has been reported for
electron gases \cite{Electron}, acoustic waves \cite{Sound}, light
waves \cite{Maret,Light}, and matter waves in a controlled
disorder \cite{BE}. It is generally agreed that the phenomena of
Anderson localization are based on interference between multiple
scattering paths, leading to localized wave functions with
exponentially decaying profiles and dense eigenspectrum
\cite{And,And+}. Theoretically, nonlinear Schr\"odinger models
offer a mean-field approximation, where the nonlinear term
containing $|\psi_n|^{2s}$ absorbs the interactions between the
components of the wave field.

It has been discussed by a few authors \cite{Sh93,PS,EPL} that
NLSE with quadratic nonlinearity (i.e., $s = 1$) observes a
localization-delocalization transition above a certain critical
strength of nonlinear interaction. That means that the localized
state is destroyed, and the nonlinear field can spread across the
lattice despite the underlying disorder, provided just that the
$\beta$ value exceeds a maximal allowed value. Below the
delocalization border, the field is dynamically localized
similarly to the linear case.

A generalization of this result to super-quadratic nonlinearity,
with $s > 1$, is far from trivial. In a recent investigation of
NLSE with disorder, we have shown \cite{PRE14} that the critical
strength destroying localization is only preserved through
dynamics, if $s = 1$. For $s > 1$ (and similarly for $0 < s < 1$,
a regime not considered here), the critical strength is dynamic in
that it involves a dependence on the number of already excited
modes (the latter are the exponentially localized modes of the
linear disordered lattice). If the field is spread across $\Delta
n$ states, then the conservation of the probability implies that
$|\psi_n|^{2} \propto 1/ \Delta n$. As the number of already
excited modes is proportional to $\Delta n$, the distance between
the frequencies obeys $\delta\omega \propto 1 / \Delta n$; whereas
the nonlinear frequency shift varies as $\Delta\omega_{\rm NL}
\propto 1/ (\Delta n)^s$. Hence $\delta\omega / \Delta\omega_{\rm
NL} \propto (\Delta n)^{s-1}$ is only independent of $\Delta n$,
if the nonlinearity is quadratic, i.e., $s=1$. The implication is
that the effect of quadratic nonlinearity does not depend on the
range of field distribution; but the effect of super-quadratic (as
well as sub-quadratic, with $0 < s < 1$) power nonlinearity does.
Hence, if initial behavior is chaotic, say, chaos remains while
spreading only for $s=1$. For $s > 1$, a transition to regularity
occurs, which blocks spreading in vicinity of the criticality
beyond a certain limiting number of excited modes $\Delta
n_{\max}$ ($\Delta n_{\max} \gg 1$).

One sees that quadratic nonlinearity, characterized by $s=1$,
plays a dynamically very distinguished role in that it is the only
type of power nonlinearity permitting an abrupt
localization-delocalization transition with unlimited spreading of
the wave field already at the delocalization border. This
localization-delocalization transition bears signatures, enabling
to associate it with a percolation transition on the infinite
Cayley tree (Bethe lattice) \cite{EPL,PRE14}. The main idea here
is that delocalization occurs through infinite clusters of chaotic
states on a Bethe lattice, with occupancy probabilities decided by
the strength of nonlinear interaction. Then the percolation
transition threshold can be translated into a critical value of
the nonlinearity control parameter, such that above this value the
field spreads to infinity, and is dynamically localized in spite
of these nonlinearities otherwise. This critical value when
account is taken for hierarchical geometry of the Cayley tree is
found to be $\beta_c = 1/\ln 2 \approx 1.4427$ \cite{EPL,PRE14}, a
fancy number representing the topology of nonlinear interaction
posed by the quadratic power term. It was argued based on a random
walk approach that in vicinity of the criticality the spreading of
the wave field is subdiffusive in the limit $t\rightarrow+\infty$,
and that the second moments grow with time as a power law
\begin{equation}
\langle (\Delta n) ^2 (t) \rangle \propto t^{\alpha}, \ \ \ t\rightarrow+\infty,
\label{MS} 
\end{equation}
with $\alpha = 1/3$ exactly. This critical regime is modeled as a
next-neighbor random walk at the onset of percolation on a Cayley
tree. The phenomena of critical spreading find their significance
in some connection with the general problem \cite{ChV} of
transport along separatrices of dynamical systems with many
degrees of freedom and are mathematically related to a description
\cite{PRE09,NJPa,NJPb,Chapter} in terms of Hamiltonian pseudochaos
(random non-chaotic dynamics with zero Lyapunov exponents)
\cite{JMPB,Report} and time-fractional diffusion equations.

For $s > 1$, the phenomena of field spreading are limited to
finite clusters at the onset of delocalization \cite{PRE14}.
Mathematically, this regime of field spreading is complicated by
the fact that finiteness of clusters on which the transport
processes concentrate conflicts with the assumptions of threshold
percolation, breaking the universal scaling laws
\cite{Havlin1,Havlin2,Naka,UFN} which pertain to the infinite
clusters. It is not clear, therefore, how to predict and explain
transport on finite clusters, avoiding as the conceptual key
element the use of percolation, and what the ensuing transport
laws would be. The goal of the present study is to present a
general solution to this problem.

The approach, which we advocate, is based on topological methods
and in a sketchy form comprises three basic steps explored in Sec.
II:

Step 1: Enabling an equivalent reduced dynamical model of
field-spreading based on backbone map.

Step 2: Projecting dynamical equations on a Cayley tree with
appropriately large coordination number which accommodates the
power nonlinearity $s$ ($s\geq 1$).

Step 3: Calculating the index of anomalous diffusion based on
combinatorial arguments, using a triangulation procedure in the
mapping space and the notion of one-bond-connected (OBC)
polyhedron.

It is shown in Sec. III that the transport on finite clusters is
subdiffusive with a power law memory kernel (for time scales for
which the dynamics concentrate on a self-similar geometry) and
pertains to a class of non-Markovian transport processes described
by generalized diffusion equations with the fractional derivative
in time. We summarize our findings in Sec. IV.

\section{The three-step topological approach}

Expanding $\psi_n$ over a basis of linearly localized modes, the
eigenfunctions of the linear problem, $\{\phi_{n,m}\}$, $m=
1,2,\dots$, we write, with time depending complex coefficients
$\sigma_m (t)$,
\begin{equation}
\psi_n = \sum_m \sigma_m (t) \phi_{n,m}.
\label{3} 
\end{equation}
We consider $\psi_n$, $\psi_n\in\{\psi_n\}$, as a vector in
functional space whose basis vectors $\phi_{n,m}$ are the Anderson
eigenstates. For strong disorder, dimensionality of this space is
infinite (countable). It is convenient to think of  each node $n$
as comprising a countable number of ``compactified" dimensions
representing the components of the wave field. So these hidden
dimensions when account is taken for Eq.~(\ref{3}) are ``expanded"
via a topological mapping procedure to form the functional space
$\{\psi_n\}$. We consider this space as providing the embedding
space for dynamics. Further, given any two vectors
$\psi_n\in\{\psi_n\}$ and $\phi_n\in\{\psi_n\}$, we define the
inner product, $\langle\psi_{n} \circ \varphi_{n}\rangle$,
\begin{equation}
\langle\psi_{n} \circ \varphi_{n}\rangle = \sum_n \psi_n^* \varphi _{n},
\label{Inn} 
\end{equation}
where star denotes complex conjugate. To this end, the functional
space $\{\psi_n\}$ becomes a Hilbert space, permitting the notions
of length, angle, and orthogonality by standard methods
\cite{Hirsch}. With these implications in mind, we consider the
functions $\phi_{n,m}$ as ``orthogonal" basis vectors obeying
\begin{equation}
\sum _n \phi^*_{n,m}\phi_{n,k} = \delta_{m,k},
\label{Kro} 
\end{equation}
where $\delta_{m,k}$ is Kronecker's delta. Then the total
probability being equal to 1 implies
\begin{equation}
\langle\psi_n\circ\psi_n\rangle =
\sum_n \psi_n^*\psi_n = \sum_m \sigma_m^* (t)\sigma_m (t) = 1.
\label{Tot} 
\end{equation}

\subsection{Step 1: The backbone map}

We define the power $2s$ ($2s\geq 2$) of the modulus of the wave
field as the power $s$ of the probability density, i.e.,
$|\psi_n|^{2s} \equiv \left[\psi_n\psi_n^*\right]^s$. Then in the
basis of linear localized modes we can write, with the use of
$\psi_n = \sum_m \sigma_m \phi_{n,m}$,
\begin{equation}
|\psi_n|^{2s} = \left[\sum_{m_1,m_2}
\sigma_{m_1} \sigma^*_{m_2} \phi_{n,m_1}\phi^*_{n,m_2}\right]^s.
\label{2s} 
\end{equation}
It is convenient to consider the expression on
the right-hand side as a functional map
\begin{equation}
\hat \mathrm{F}_s:\{\phi_{n,m}\}\rightarrow
\left[\sum_{m_1,m_2} \sigma_{m_1} \sigma^*_{m_2} \phi_{n,m_1}\phi^*_{n,m_2}\right]^s
\label{3s} 
\end{equation}
from the vector field $\{\phi_{n,m}\}$ into the scalar field
$|\psi_n|^{2s}$. It is noticed that the map in Eq.~(\ref{3s}) is
positive definite, and that it contains a self-similarity
character in it, such that by stretching the basis vectors (by a
stretch factor $\lambda$) the value of $\hat \mathrm{F}_s$ is just
renormalized (multiplied by $|\lambda|^{2s}$). We have,
accordingly,
\begin{equation}
\hat \mathrm{F}_s\{\lambda \phi_{n,m}\} =
|\lambda|^{2s} \hat \mathrm{F}_s\{\phi_{n,m}\}.
\label{4s} 
\end{equation}
Consider expanding the power law on the right-hand side of
Eq.~(\ref{2s}). If $s$ is a positive integer, then a regular
expansion can be obtained as a sum over $s$ pairs of indices
$(m_{1,1}, m_{1,2})\dots (m_{s,1}, m_{s,2})$. The result is a
homogeneous polynomial, an $s$-quadratic form \cite{Cox}. In
contrast, for fractional $s$, a simple procedure does not exist.
Even so, with the aid of Eq.~(\ref{4s}), one might circumvent the
problem by proposing that the expansion goes as a homogeneous
polynomial whose nonzero terms all have the same degree $2s$.
``Homogeneous" means that every term in the series is in some
sense representative of the whole. Then one does not really need
to obtain a complete expansion of $\hat \mathrm{F}_s$ in order to
predict dynamical laws for the transport, since it will be
sufficient to consider a certain collection of terms which by
themselves completely characterize the algebraic structure of
$\hat \mathrm{F}_s$ as a consequence of the homogeneity property.
We dub this collection of terms the backbone, and we define it
through the homogeneous map
\begin{equation}
\hat \mathrm{F}^\prime_s:\{\phi_{n,m}\}\rightarrow
\sum_{m_1,m_2} \sigma_{m_1}^s \sigma^{*s}_{m_2} \phi^s_{n,m_1}\phi^{*s}_{n,m_2}.
\label{3ss} 
\end{equation}
In what follows, we consider the backbone as representing the
algebraic structure of $\hat \mathrm{F}_s$ in the sense of
Eq.~(\ref{4s}). So, for fractional $s$, our analysis will be based
on a reduced model which is obtained by replacing the original map
$\hat \mathrm{F}_s$ by the backbone map $\hat
\mathrm{F}^\prime_s$. The claim is that the reduction $\hat
\mathrm{F}_s\rightarrow\hat \mathrm{F}^\prime_s$ does not really
alter the scaling exponents behind the wave-spreading, since the
algebraic structure of the original map is there anyway. Note that
$\hat \mathrm{F}_s$ and $\hat \mathrm{F}^\prime_s$ both have the
same degree $2s$, which is the sum of the exponents of the
variables that appear in their terms. Note, also, that the
original map coincides with its backbone in the limit
$s\rightarrow 1$. This property illustrates the significance of
the quadratic nonlinearity {\it vs.} arbitrary power nonlinearity.
Turning to NLSE~(\ref{1s}), if we now substitute the original
power nonlinearity with the backbone map, in the orthogonal basis
of the Anderson eigenstates we find
\begin{equation}
i\dot{\sigma}_k - \omega_k \sigma_k =
\beta \sum_{m_1, m_2, m_3} V_{k, m_1, m_2, m_3} \sigma_{m_1}^s \sigma^{*s}_{m_2} \sigma_{m_3},
\label{4s+} 
\end{equation}
where
\begin{equation}
V_{k, m_1, m_2, m_3} = \sum_{n} \phi^*_{n,k}\phi_{n,m_1}^s\phi^{*s}_{n,m_2}\phi_{n,m_3}
\label{5s+} 
\end{equation}
are complex coefficients characterizing the overlap structure of
the nonlinear field, and we have reintroduced the eigenvalues of
the linear problem, $\omega_k$, satisfying $\hat H_L \phi_{n,k} =
\omega_k \phi_{n,k}$. Although obvious, it should be emphasized
that the use of the backbone map $\hat \mathrm{F}^\prime_s$ in
place of the original map $\hat \mathrm{F}_s$ preserves the
Hamiltonian character of the dynamics, but with a different
interaction Hamiltonian, $\hat H_{\rm int}$,
\begin{equation}
\hat H_{\rm int} = \frac{\beta}{1+s}
\sum_{k, m_1, m_2, m_3} V_{k, m_1, m_2, m_3}
\sigma^*_k \sigma_{m_1}^s \sigma^{*s}_{m_2} \sigma_{m_3}.
\label{6s+} 
\end{equation}
Note that $\hat H_{\rm int}$ includes self-interactions through
the diagonal elements $V_{k, k, k, k}$. Another important point
worth noting is that the strength of the interaction vanishes in
the limit $s\rightarrow\infty$ (as $\sim 1/s$). Therefore, keeping
the $\beta$ parameter finite, and letting $s\rightarrow \infty$,
one generates a regime where the nonlinear field is asymptotically
localized. One sees that high-power nonlinearities act as to
reinstall the Anderson localization. We shall confirm this by the
direct calculation of respective transport exponents.
Equations~(\ref{4s+}) define a system of coupled nonlinear
oscillators with a parametric dependence on $s$. Similarly to the
NLSE model with a quadratic power nonlinearity, each nonlinear
oscillator with the Hamiltonian
\begin{equation}
\hat h_{k} = \omega_k \sigma^*_k \sigma_k +
\frac{\beta}{1+s} V_{k, k, k, k} \sigma^*_k \sigma_{k}^s \sigma^{*s}_{k} \sigma_{k}
\label{6+h+s} 
\end{equation}
and the equation of motion
\begin{equation}
i\dot{\sigma}_k - \omega_k \sigma_k -
\beta V_{k, k, k, k} \sigma^s_{k} \sigma^{*s}_{k} \sigma_{k} = 0
\label{eq+s} 
\end{equation}
represents one nonlinear eigenstate in the system $-$ identified
by its wave number $k$, unperturbed frequency $\omega_k$, and
nonlinear frequency shift $\Delta \omega_{k} = \beta V_{k, k, k,
k} \sigma^s_{k} \sigma^{*s}_{k}$. We reiterate that non-diagonal
elements $V_{k, m_1, m_2, m_3}$ characterize couplings between
each four eigenstates with wave numbers $k$, $m_1$, $m_2$, and
$m_3$. The comprehension of Hamiltonian character of the dynamics
paves the way for a consistency analysis of the various transport
scenarios behind the Anderson localization problem (with the
topology of resonance overlap taken into account)
\cite{EPL,PRE14}. To this end, the transport problem for the wave
function becomes essentially a topological problem in phase space.

\subsection{Step 2: Mapping on a Cayley tree}

The ``edge" character of onset transport corresponds to infinite
chains of next-neighbor interactions with a minimized number of
links at every step. For the reasons of symmetry, when summing on
the right-hand side of Eq.~(\ref{4s+}), the only combinations of
terms to be taken into account, apart from the self-interaction
term $\sigma_{k}^s \sigma^{*s}_{k} \sigma_{k}$, are, essentially,
$\sigma_{k-1}^s \sigma^{*s}_{k} \sigma_{k+1}$ and $\sigma_{k+1}^s
\sigma^{*s}_{k} \sigma_{k-1}$. These terms will come with
respective interaction amplitudes $V_{k, k, k, k}$, $V_{k, k-1, k,
k+1}$, and $V_{k, k+1, k, k-1}$, which we shall denote simply by
$V_k$, $V_k^-$, and $V_k^+$. Then on the right-hand side (r.h.s.)
of Eq.~(\ref{4s+}) we have
\begin{equation}
{\rm r.h.s.} = \beta V_{k} \sigma_{k}^s \sigma^{*s}_{k} \sigma_{k} +
\beta \sum_{\pm}V_{k}^\pm \sigma_{k\pm1}^s \sigma^{*s}_{k} \sigma_{k\mp1}.
\label{4s++} 
\end{equation}
The interaction Hamiltonian in Eq.~(\ref{6s+}) becomes
\begin{equation}
\hat H_{\rm int} = \frac{\beta}{1+s} \sum_{k}
\left[V_{k} \sigma^*_k \sigma_{k}^s \sigma^{*s}_{k} \sigma_{k} +
\sum_{\pm}V_{k}^\pm \sigma^*_k \sigma_{k\pm1}^s \sigma^{*s}_{k} \sigma_{k\mp1}\right]
\label{6s++} 
\end{equation}
representing the effective reduced $\hat H_{\rm int}$ for
arbitrary real power $s\geq 1$. Assuming that the exponent $s$ is
confined between two integer numbers, i.e., $j\leq s < j+1$, in
the next-neighbor interaction term we can write
\begin{equation}
\hat H^\prime_{\rm int} = \frac{\beta}{1+s} \sum_{k} \sum_{\pm}V_{k}^\pm
\left[\sigma^*_k \sigma_{k\pm1}^j \sigma^{*j}_{k} \sigma_{k\mp1}\right]
\sigma_{k\pm1}^{s-j} \sigma^{*s-j}_{k},
\label{6s++p} 
\end{equation}
where the prime symbol indicates that we have extracted the
self-interactions. When drawn on a graph in wave-number space, the
terms raised to the power $s-j$ will correspond to disconnected
bonds, thought as Cantor sets with the fractal dimensionality
$0\leq s-j < 1$. Hence, they will not contribute to
field-spreading. These terms, therefore, can be cut off from the
interaction Hamiltonian, suggesting that only those terms raised
to the integer power, $j$, should be considered. We have,
accordingly,
\begin{equation}
\hat H^\prime_{\rm int} \rightarrow \frac{\beta}{1+s}
\sum_{k} \sum_{\pm}V_{k}^\pm \sigma^*_k \sigma_{k\pm1}^j \sigma^{*j}_{k} \sigma_{k\mp1}.
\label{6s++c} 
\end{equation}
This is the desired result. Equation~(\ref{6s++c}) defines the
effective reduced interaction Hamiltonian in the parameter range
of onset spreading for $j\leq s < j+1$.

Focusing on the transport problem for the wave field,
because the interactions are next-neighbor-like,
it is convenient to project the system of coupled
dynamical equations~(\ref{4s++}) on a Cayley tree,
such that each node with the coordinate $k$ represents
a nonlinear eigenstate, or nonlinear oscillator with the equation of
motion~(\ref{eq+s}); the outgoing bonds represent the complex
amplitudes $\sigma_{k\pm 1}$ and $\sigma_{k\mp 1}$; and
the ingoing bonds, which involve complex conjugation,
represent the complex amplitudes $\sigma_{k}^*$.
To make it with the amplitudes $\sigma_{k}^*$
when raised to the algebraic power $s$ one needs for each
node a fractional number $s$ of the ingoing bonds.
Confining the $s$ value between two nearest integer numbers,
$j\leq s < j+1$, we carry on with $j$ connected bonds,
which we charge to receive the interactions, and one
disconnected bond, which corresponds to a Cantor set
with the fractal dimensionality $s-j$, and which cannot transmit the waves.
At this point we cut this bond off the tree. A similar procedure applied
to the amplitudes $\sigma_{k\pm 1}$, coming up in the algebraic power $s$,
generates $j$ outgoing bonds, leaving one disconnected bond behind.
Lastly, the remaining amplitude $\sigma_{k\mp 1}$, which does not
involve a nonlinear power, contributes with one outgoing bond for
each combination of the indexes. One sees that the mapping requires a
Cayley tree with the coordination number $z = 2j+1$.

\subsection{Step 3: Obtaining the connectivity index}

\begin{figure}[t]
\includegraphics[width=0.7\hsize]{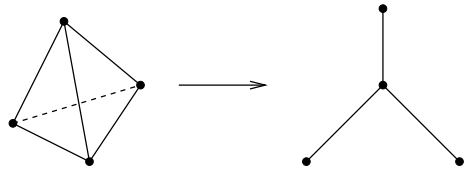}
\caption{Determination of one-bond-connections for a tetrahedron
in $R^3$.} \label{draw1}
\end{figure}

If the interactions are next-neighbor-like, and if the number of
excited modes after $t$ time steps is $\Delta n (t)$, then
self-similarity will imply that
\begin{equation}
\langle (\Delta n) ^2 (t) \rangle \propto t^{2 / (2+\theta)}, \ \ \ t\rightarrow+\infty,
\label{MS+} 
\end{equation}
where $\theta$ is the connectivity exponent of the structure on
which the spreading processes occur. This exponent accounts for
the deviation from the usual Fickian diffusion in a self-similar
geometry \cite{Havlin1,Havlin2,Naka,Gefen} and observes remarkable
invariance properties under homeomorphic maps of fractals
\cite{UFN,PRE97}. The scaling law in Eq.~(\ref{MS+}) has been
discussed by Gefen et al. \cite{Gefen} for anomalous diffusion on
percolation clusters. The crucial assumption behind this scaling,
however, is the assumption of self-similarity (and {\it not} of
percolation) extending the range of validity of Eq.~(\ref{MS+}) to
any self-similar fractal. Here we apply the scaling law in
Eq.~(\ref{MS+}) to Cayley trees by appropriately choosing the
$\theta$ value. We note in passing that self-similarity of the
Cayley trees is not necessarily manifest in their folding in the
embedding space, but is inherent in their connectedness and
topology \cite{Schroeder}. Indeed a Cayley tree is a graph without
loops, where each node hosts the same number of branches (known as
the coordination number). Therefore, one might expect from the
outset that the value of $\theta$ will be a function of the
coordination number, given that the dynamics occur on a Cayley
tree. The coordination number, in its turn, will depend on the
power $s$, thus paving the way to predict onset spreading in
association with the topology of interaction between the
components of the wave field.

More so, the power nonlinearity in Eq.~(\ref{1s}) suggests that
the connectivity value is a {\it multiplicative} function of $s$.
Also one might expect this function to naturally reproduce the
known value $\theta = 4$ \cite{Havlin1,Havlin2,Coniglio} for
mean-field percolation on Bethe lattices in the limit
$s\rightarrow 1$. Then the obvious dependence satisfying these
criteria is $\theta = 4s$, where $s\geq 1$. For the integer and
half-integer $s$, this dependence can also be derived using the
standard renormalization-group procedure \cite{Shaugh} for
self-similar clusters in Hilbert space. So restricting ourselves
to the short times for which the dynamics concentrate on a
self-similar geometry, we write, with $\Delta n_{\max} \gg 1$,
\begin{equation}
\langle (\Delta n) ^2 (t) \rangle \propto t^{1 / (2s+1)},
\ \ \ 1\ll t\ll (\Delta n_{\max})^{2(2s+1)},
\label{Finite} 
\end{equation}
from which the scaling dependence in Eq.~(\ref{MS}) can be deduced
for quadratic nonlinearity, i.e., $\alpha = 1/3$. Numerically, the
field-spreading on finite clusters has been already discussed
\cite{Iomin,Skokos_PRE} based on computer simulation results,
using one-dimensional disordered Klein-Gordon chains with tunable
nonlinearity. It is noticed that the exponent of the powerlaw,
$\alpha = 1/ (2s+1)$, vanishes in the limit $s\rightarrow \infty$,
conformally with the previous considerations.

To illustrate the determination of $\theta$ and to address the
origin of subdiffusion in the regime of next-neighbor
communication rule, we look directly into the connectivity
properties of finite clusters. For this, we need a simple
procedure by which calculations can be done exactly. We formulate
such a procedure for integer and half-integer values of $s$ using
topological triangulation \cite{Sen,Fom} of the Cayley tree. For
the purpose of formal analysis, it is essential to choose a node
on a Cayley tree and a reference system of $z$ next-neighbor
connecting bonds (for a Cayley tree with the coordination number
$z = 2s +1$). Next we dispose the selected node of the Cayley tree
and immerse it into a $z$-dimensional Euclidean space $R^{2s+1}$.
The latter space is built on $2s+1$ orthonormal basis vectors.
Note that there is a one-to-one correspondence between the basis
vectors in $R^{2s+1}$ and the reference bonds on the Cayley tree.
Connecting the ending points of the basis vectors generates a
polyhedron in $R^{2s+1}$, which reflects the connectivity of the
original Cayley tree and the hierarchical composition of this. For
the standard Cayley tree with $z=3$ the associated geometric
construction is illustrated in Fig.~1.

More so, we apply the above procedure to {\it all} nodes of the
original Cayley tree, such that the nodes which communicate via a
next-neighbor rule on the tree go to the ending points of the
corresponding basis vectors in $R^{2s+1}$. One sees that this
procedure generates an infinite chain of mutually overlapping
polyhedrons. The number of internal one-bond-connections (OBC's)
is obtained as the {\it minimal} number of bonds belonging to the
same polyhedron and enabling an infinite connected mesh. We
distinguish between ``nodes" which compose a polyhedron, that is
analyzed, and ``node-vertexes" which are nodes pertaining to
neighboring polyhedrons. In what follows, we identify the nodes
with numbers, and node-vertexes with letters. For instance, the
only node having a full family of nearest neighbors in Figs.~2
and~3 is the node marked as 1. The connectivity index $\theta$ is
obtained as the number of paths (routes without self-crossings)
connecting node 1 to node $2s+2$ via any nodes of the {\it same}
polyhedron.

\begin{figure}[t]
\includegraphics[width=0.7\hsize]{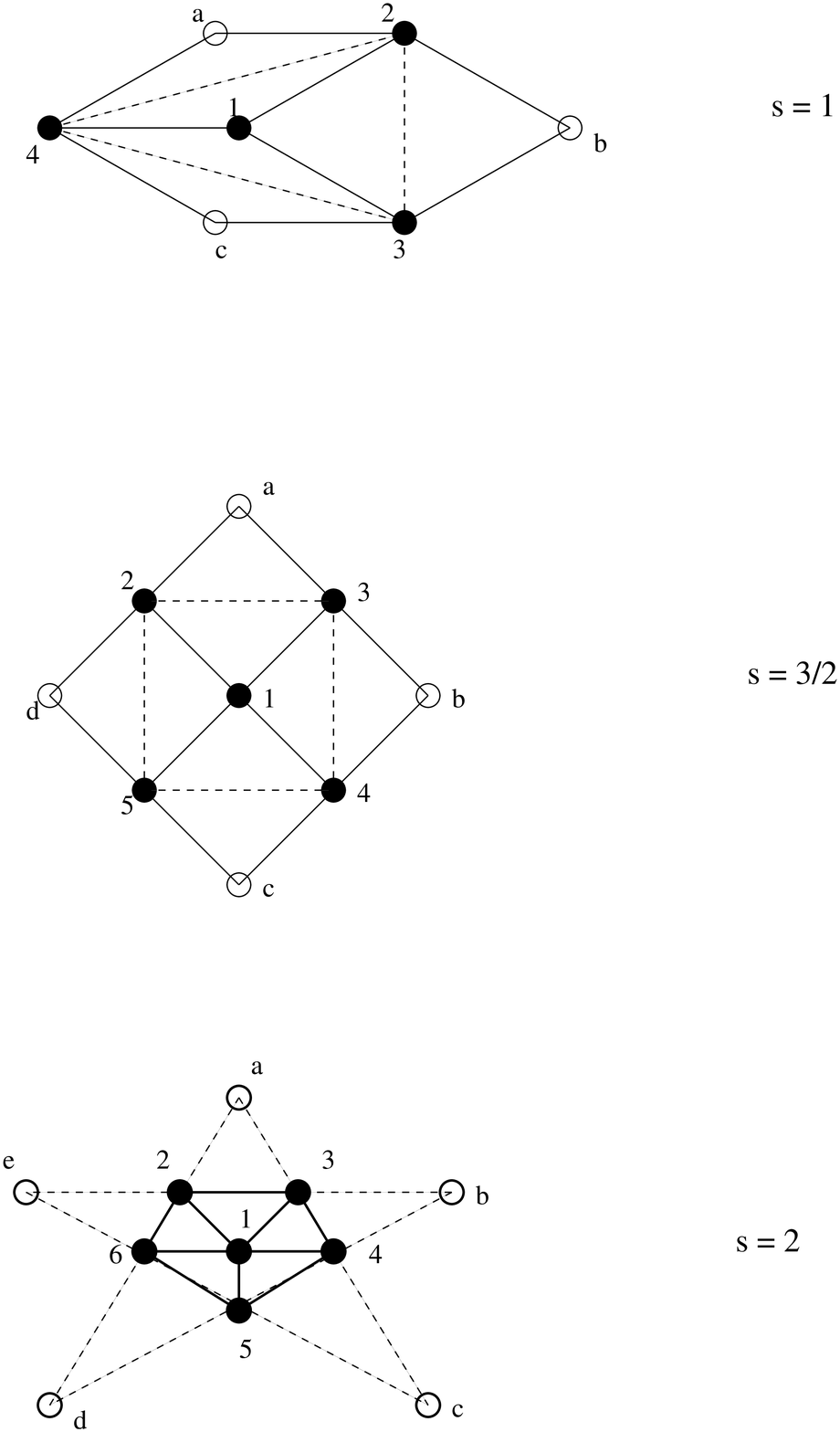}
\caption{Schematic representation of polyhedrons for $s=1$, $3/2$,
and $2$. Nodes belonging to the same polyhedron are marked by
numbers: $1$, $2$, $3$, etc. Nodes-vertexes pertaining to other
polyhedrons are marked by letters: $a$, $b$, $c$, etc.
One-bond-connections are shown by solid lines; virtual connections
between the nodes, by broken lines.} \label{draw2}
\end{figure}
\begin{figure}[t]
\includegraphics[width=0.7\hsize]{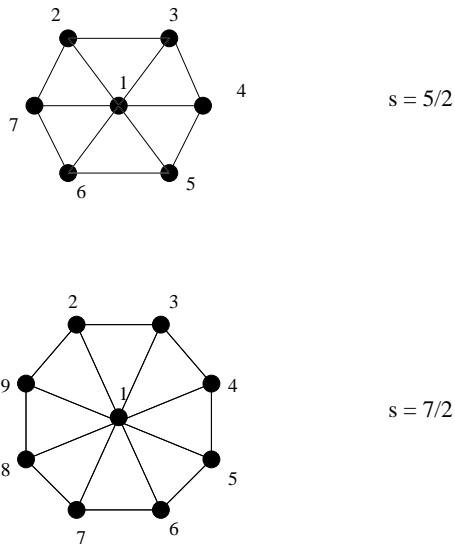}
\caption{Schematic representation of polyhedrons for the
half-integer $s=5/2$ and $s=7/2$.} \label{draw3}
\end{figure}

To illustrate, consider quadratic nonlinearity first, with $s = 1$
(see Fig.~2). Clearly, there are just three OBC's defined by a
tetrahedron with nodes $1, 2, 3$ and $4$. So one identifies these
OBC's with the bonds $(1-2)$, $(1-3)$, and $(1-4)$. It is noticed
that the connection between nodes $3$ and $4$ occurs via the
node-vertex $c$; the connection between nodes $2$ and $3$ occurs
via the node-vertex $b$; and the connection between nodes $2$ and
$4$, via the node-vertex $a$. The connectivity index $\theta$ is
the number of paths connecting node $1$ to node $2s + 2 = 4$.
These paths are just four, namely, $(1-4)$, $(1-2-4)$, $(1-3-4)$,
and $(1-2-3-4)$. Hence, $\theta = 4$. This result is to be
expected, as it also characterizes mean-field transport on lattice
animals \cite{Havlin1,Havlin2} and trees \cite{Coniglio}. With the
aid of Eqs.~(\ref{MS+}) and~(\ref{Finite}) one also obtains
$\langle (\Delta n) ^2 (t) \rangle \propto t^{1/3}$ for $t\ll
(\Delta n_{\max})^{6}$ consistently with the result of Ref.
\cite{PRE14}.

Let us now calculate the connectivity index $\theta$ for
half-integer $s=3/2$. Here one constructs a pentahedron in
$R^{4}$, with nodes marked $1, 2, 3, 4$ and $5$. The OBC's are the
bonds $(1-2)$, $(1-3)$, $(1-4)$, and $(1-5)$ (see Fig.~2). There
are exactly six paths connecting node 1 to node 5, that is,
$(1-5)$, $(1-2-5)$, $(1-4-5)$, $(1-3-2-5)$, $(1-3-4-5)$, and
$(1-2-3-4-5)$. Thus, $\theta = 6$, leading to a subdiffusive
scaling of second moments $\langle (\Delta n) ^2 (t) \rangle
\propto t^{1/4}$ for $t\ll (\Delta n_{\max})^{8}$.

The same triangulation procedure applied to a hexahedron in
$R^{5}$ generates for $s=2$ the following eight paths (see
Fig.~2): $(1-6)$, $(1-2-6)$, $(1-5-6)$, $(1-3-2-6)$, $(1-4-5-6)$,
$(1-4-3-2-6)$, $(1-3-4-5-6)$, and $(1-2-3-4-5-6)$, leading to
$\theta = 8$. The scaling of second moments is given by $\langle
(\Delta n) ^2 (t) \rangle \propto t^{1/5}$ for $t\ll (\Delta
n_{\max})^{10}$.

In Fig.~3 we also present for reader's convenience respective
geometric constructions corresponding to half-integer $s = 5/2$
and $s = 7/2$, facilitating the calculation of the paths and of
respective connectivity values.

By mathematical induction the connectivity index for integer and
half-integer $s$ is given by $\theta = 4s$, yielding for the
transport exponent $\alpha = 2 / (2+\theta) = 1 / (2s+1)$
consistently with the subdiffusive scaling in Eq.~(\ref{Finite}).
Eliminating $s$ with the aid of coordination number, $z = 2s +1$,
we also get $\alpha = 1 / z$. One sees that the transport is
slowed down by complexity elements of clusters, contained in the
$z$ value. All in all, one sees that higher-order nonlinearities
($s > 1$) have a progressively weakening effect over the transport
rates, with the fastest transport obtained for quadratic power
nonlinearity.

\section{Non-Markovian diffusion equation}

The next-neighbor communication rule which we associate with the
phenomena of onset spreading must have implications for anomalous
diffusion on the short times for which the dynamics concentrate on
a self-similar geometry of finite clusters. This equation has been
already discussed \cite{Chapter,PRE09,NJPa,NJPb} and has been
shown to be a non-Markovian variant of the diffusion equation with
powerlaw memory kernel:
\begin{equation}
\frac{\partial}{\partial t} f (t, \Delta n) =
\frac{\partial}{\partial t}\int _{0}^{t} \frac{dt^{\prime}}{(t - t^{\prime})^{1-\alpha}}
\frac{\partial^2}{\partial (\Delta n)^2} \left[W_\theta f (t^\prime,
\Delta n)\right],\label{23} 
\end{equation}
where $1-\alpha = \theta / (2 + \theta)$; $\theta$ is the
connectivity exponent; $W_\theta$ absorbs the parameters of the
transport model; and we have chosen $t=0$ as the beginning of the
system's time evolution. The integral term on the right-hand side
has the analytical structure of fractional time the so-called
Riemann-Liouville fractional derivative \cite{Podlubny}. In a
compact form,
\begin{equation}
\frac{\partial}{\partial t} f (t, \Delta n) =
\frac{\partial^{1-\alpha}}{\partial t^{1-\alpha}}
\frac{\partial^2}{\partial (\Delta n)^2} \left[W_\theta f (t, \Delta n)\right].\label{24}
\end{equation}
The fractional order of time differentiation in Eq.~(\ref{24}) is
determined by the connectivity value through $1-\alpha = \theta /
(2+\theta)$ and is exactly zero for $\theta = 0$. Then the
fractional derivative of the zero order is a unity operator,
implying that no fractional properties come into play for
homogeneous spaces. Also in writing Eq.~(\ref{23}) we have adopted
results of Refs.~\cite{PRE09,NJPa,NJPb} to diffusion processes on
a single cluster. Equations~(\ref{23}) and~(\ref{24}) when account
is taken for the initial value problem can be rephrased
\cite{PRE09} in terms of the Caputo fractional derivative
\cite{Podlubny} which shows a better behavior under
transformations.

One sees that the dispersion law in Eq.~(\ref{MS+}) can be
obtained as a second moment of the fractional diffusion
equation~(\ref{23}), with $\alpha = 2 / (2+\theta)$. Using for the
connectivity exponent $\theta = 4s$, one also finds the fractional
order $1-\alpha = 2s / (2s + 1)$ in the entire parameter range $s
\geq 1$, showing that ordinary differentiation is reinstalled on
the right-hand side of Eq.~(\ref{24}) in the limit
$s\rightarrow\infty$. For $s=1$, one gets $1-\alpha = 2/3$,
implying that the diffusion process is essentially non-Markovian
with power-law correlations in the regime of quadratic
nonlinearity. We associate this non-Markovian character of
field-spreading with the effect of complexity elements of the
Cayley tree, contained in the $z=2s+1$ value.

It is noticed that the fractional diffusion equation in
Eq.~(\ref{24}) is ``born" within the exact mathematical framework
of nonlinear Schr\"odinger equation with usual time
differentiation. Indeed, no {\it ad hoc} introduction of
fractional time differentiation in the dynamic Eq.~(\ref{1s}) has
been assumed to obtain this sub\-diffusion. It is, in fact, the
interplay between nonlinearity and randomness, which leads to a
non-Markovian transport of the wave function at criticality, and
to a time-fractional kinetic equation in the end. This observation
also emphasizes the different physics implications behind the
fractional kinetic {\it vs.} dynamical equations
\cite{Chapter,IominFT,Comb}. Equation~(\ref{24}) shows that the
onset spreading is a matter of {fractional}, or ``strange,"
kinetics \cite{Klafter,Nature1,Nature2} consistently with the
implication of critical behavior \cite{PRE09,Chapter,Report}.

The fundamental solution or Green's function of the fractional
Eq.~(\ref{24}) is evidenced in Table~1 of Ref.~\cite{Rest}.

\section{Conclusions}

This study is concerned with destruction of Anderson localization
by a nonlinearity of the power-law type. It has been proposed
using an NLSE with random potential on a lattice that quadratic
nonlinearity plays a dynamically very distinguished role in that
it is the only type of power nonlinearity permitting an abrupt
localization-delocalization transition with unlimited spreading
already at the delocalization border. For super-quadratic
nonlinearity the borderline spreading corresponds to a diffusion
process on finite clusters. We have suggested an analytical method
to predict and explain such transport processes. Our method uses a
topological approximation of the nonlinear Anderson model and, if
the exponent of the power nonlinearity is either integer or
half-integer, will yield the wanted value of the transport
exponent via a triangulation procedure in an Euclidean mapping
space. Also we predict that the transport of waves at the border
of delocalization is subdiffusive, with the exponent $\alpha$
which is inversely proportional with the power nonlinearity
increased by one. For quadratic nonlinearity we have $\langle
(\Delta n) ^2 (t) \rangle \propto t^{1/3}$ for $t\rightarrow
+\infty$ consistently with the previous investigations
\cite{EPL,PRE14}. A kinetic picture of the transport arising from
these investigations uses a fractional extension of the diffusion
equation to fractional derivatives over the time, signifying
non-Markovian dynamics with algebraically decaying time
correlations.

\acknowledgments A.V.M. and A.I. thank the Max-Planck-Institute
for the Physics of Complex Systems for hospitality and financial
support. This work was supported in part by the Israel Science
Foundation (ISF) and by the ISSI project ``Self-Organized
Criticality and Turbulence" (Bern, Switzerland).

\end{document}